\documentclass{PoS}

\title{Meson spectrum of Sp(4) lattice gauge theory with two fundamental Dirac fermions}

\ShortTitle{Mesons in two-flavour Sp(4)}

\author{\speaker{Jong-Wan Lee},$^{ab}$\thanks{Supported in part by the National Research Foundation of Korea grant funded by the Korea government(MSIT) (NRF-2018R1C1B3001379) and in part by Korea Research Fellowship programme funded by the Ministry of Science, ICT and Future Planning through the National Research Foundation of Korea (2016H1D3A1909283).} 
	     Ed Bennett,$^c$\thanks{Funded in part by the Supercomputing Wales project, which is part-funded by the European Regional Development Fund (ERDF) via Welsh Government.} 
	     Deog Ki Hong,$^a$\thanks{Supported by Basic Science Research Program through the National Research Foundation of Korea funded by the Ministry of Education (NRF-2017R1D1A1B06033701).}
	     C.-J.~David Lin,$^{de}$\thanks{Supported by Taiwanese MoST grant 105-2628-M-009-003-MY4.} 
	     Biagio Lucini,$^f$\thanks{Supported in part by the Royal Society Wolfson Research Merit Award WM170010.} \footnotemark[7]
	     Maurizio Piai,$^g$\thanks{Supported in part by the STFC Consolidated Grants ST/L000369/1 and ST/P00055X/1.} 
             and Davide Vadacchino$^{h}$\\
             \llap{$^a$}Department of Physics, Pusan National University, 
			Busan 46241, Korea\\
             \llap{$^b$}Extreme Physics Institute, Pusan National University, 
			Busan 46241, Korea\\ 
             \llap{$^c$}Swansea Academy of Advanced Computing, Bay Campus, Swansea University, 
		        Swansea, SA1 8EN, UK\\
             \llap{$^d$}Institute of Physics, National Chiao-Tung University, 
			Hsinchu 30010, Taiwan\\
             \llap{$^e$}Centre for High Energy Physics, Chung-Yuan Christian University, 
			Chung-Li 32032, Taiwan\\
             \llap{$^f$}Department of Mathematics, Computational Foundary, Bay Campus, Swansea University, 
			Swansea, SA1 8EN, UK\\
             \llap{$^g$}Department of Physics, Swansea University, 
			Singleton Park, Swansea, SA2 8PP, UK\\
	     \llap{$^h$}INFN, Sezione di Pisa, 
			Largo Pontecorvo 3, 56127 Pisa, Italy\\
             E-mail:  \email{jwlee823@pusan.ac.kr}, 
		      \email{e.j.bennett@swansea.ac.uk}, 
		      \email{dkhong@pusan.ac.kr}, 
		      \email{dlin@mail.nctu.edu.tw}, 
		      \email{b.lucini@swansea.ac.uk},
		      \email{m.piai@swansea.ac.uk}, 
		      \email{davide.vadacchino@pi.infn.it}}

\abstract{
We calculate the meson spectrum of the 
 $Sp(4)$ lattice gauge theory coupled to two fundamental flavours of dynamical Dirac fermions.
We focus on some of the lightest (flavoured) spin-0 and spin-1 states. 
This theory provides an ultraviolet completion for composite Higgs models based upon the 
$SU(4)/Sp(4)$ coset.
We analyse the  strongly coupled dynamics in isolation, without explicit coupling to the standard model. 
We carry out continuum extrapolations using dynamical ensembles generated at five different 
values of bare lattice coupling,
 and for several values of the bare fermion mass.
We fit the resulting meson masses and decay constants
 to a low-energy effective field theory built along the ideas of hidden local symmetry. 
We also compare our results to those of other closely related
 lattice gauge theories, which have matter content consisting of two fundamental Dirac flavours. 
}

\FullConference{37th International Symposium on Lattice Field Theory - Lattice2019\\
		16-22 June 2019\\
		Wuhan, China}

\begin{document}

\section{Introduction}

In composite Higgs models (CHMs)~\cite{Kaplan:1983fs}, 
the role of the complex Higgs doublet field of the standard model (SM) 
is played by 
pseudo-Nambu-Goldstone bosons (pNGBs)
that arise from the spontaneous breaking of a continuous global symmetry, 
within the low-energy description of a strongly-coupled new physics sector.
The Higgs potential responsible for electroweak symmetry breaking (EWSB) 
is induced dynamically by weakly coupling the pNGBs to the SM gauge bosons and fermions. 
Furthermore, the relatively large mass of the top quark
can also be accommodated by supplementing this construction
by the idea of partial compositeness~\cite{Kaplan:1991dc}. 
Other heavy composite states, besides the pNGBs and top partners, 
emerge from the new physics sector.  
Quantitative studies of such heavy resonances provide useful input both
for model building and for collider searches.

Developing the research programme announced in Ref.~\cite{Bennett:2017kga},
in this contribution we consider the  $Sp(4)$ lattice gauge
 theory with $N_f=2$ fundamental Dirac fermions. 
This theory provides a dynamical origin for CHMs based upon the $SU(4)/Sp(4)$ coset.  
We adopt the Wilson lattice formulation of the dynamical fermions,
 implement  the hybrid Monte Carlo (HMC) method, 
and extract masses and decay constants 
of  spin-0 and spin-1 (flavoured) mesons  from the
relevant 2-point  correlation functions in Euclidean space.
 
We repeat our calculations for several values of bare lattice parameters and,
 borrowing the ideas of Wilson chiral perturbation theory (W$\chi$PT), 
we carry out the first continuum and massless extrapolations for this theory. 
We then restrict our attention to the lightest pseudoscalar, vector and axial vector particles. 
We analyse the continuum extrapolated data by means of a low-energy effective field theory (EFT), 
that extends the continuum chiral Lagrangian by implementing
the principles of hidden local symmetry (HLS). 
The model studied in this work is also relevant in the context of strongly
 interacting massive particles (SIMP) as dark matter candidates~\cite{Hochberg:2014kqa}. 
Our quantitative results on the low-energy spectrum are useful input in this context.
Complete details of this study can be found in Ref.~\cite{Bennett:2019jzz,quenched}.

\section{Lattice theory}

The four-dimensional Euclidean  (unimproved) lattice action, for the plaquette $P_{\mu\nu}$
 and the massive Wilson-Dirac fermions $Q$, is given by
\begin{equation}
S\equiv \beta \sum_x \sum_{\mu < v}\left(1-\frac{1}{4} {\rm Re} \, {\rm Tr} \, P_{\mu\nu}(x)\right)
+a^4\sum_x \bar{Q}^i(x) (D+\,m_0) Q^i(x),
\end{equation}
where $\beta=8/g^2$ is the bare lattice coupling, $a$ is the lattice spacing, and $m_0$ is the bare fermion mass. 
The index $i=1,2$ labels the Dirac fundamental flavours. 
The plaquette $P_{\mu\nu}$ is given by $P_{\mu\nu}(x)=U_\mu(x) U_\nu (x+\hat{\mu}) U_\mu^\dagger(x+\hat{\nu}) U_\nu^\dagger (x)$, 
while the massless Dirac operator $D$ is defined as
\begin{equation}
D Q(x) \equiv \frac{4}{a} Q(x)-\frac{1}{2a}\sum_\mu \left\{
(1-\gamma_\mu)U_\mu(x) Q(x+\hat{\mu})+(1+\gamma_\mu)U_\mu (x-\hat{\mu})Q(x-\hat{\mu})
\right\}. 
\end{equation}
The link variables $U_\mu(x)$ are elements of the $Sp(4)$ group,
and $\hat{\mu}$ is the unit vector in one of  the four space-time directions.
In the continuum model the global symmetry is enhanced to $SU(4)$. 
The fermion condensate breaks it spontaneously to the $Sp(4)$ subgroup,
aligned with the explicit breaking that arises both because of a degenerate fermion 
mass and of the  Wilson term at finite lattice spacing.

In our numerical studies we use a variant of the HiRep code \cite{DelDebbio:2008zf} and 
generate gauge configurations by means of the HMC algorithm. 
(For the details of the modification of the code and 
some numerical tests see~\cite{Bennett:2017kga,Bennett:2019jzz}.) 
On  four-dimensional Euclidean lattices of size $N_f\times N_s^3$ 
we impose periodic boundary conditions in all spatial directions, 
while we impose periodic and anti-periodic boundary conditions in the temporal direction 
for  gauge and  fermion fields, respectively. 

All the ensembles are generated in the weak coupling regime, 
$\beta \gtrsim 6.8$~\cite{Bennett:2017kga}, 
with the choices of $\beta=6.9, \, 7.05,\, 7.2, \, 7.4, \, 7.5$ 
(see Ref. \cite{Bennett:2019jzz} for further details). 
These ensembles satisfy also the condition $m_{\rm PS} \, L \gtrsim 7.5$, 
so that finite volume effects are statistically negligible~\cite{Bennett:2019jzz,Lee:2018ztv}. 
We adopt L\"{u}scher's gradient flow (GF) method as scale-setting scheme, 
with the Wilson flow (for the lattice definition) evolving along the fictitious time $t$ \cite{Luscher:2010iy}.
We use the particular definition of the gradient flow scale $w_0$ determined by 
$t\textrm{d}\mathcal{E}(t)/\textrm{d} t|_{t/a^2=(w_0/a)^2}=\mathcal{W}_0$ \cite{Borsanyi:2012zs}, 
where $\mathcal{E}(t)$ is the action density at non-zero flow time $t$.
The reference value is chosen to be $\mathcal{W}_0=0.35$~\cite{Bennett:2017kga}. 
Accordingly, we express all dimensional quantities in units of $w_0$,
and define  $\hat{m}=m w_0$, $\hat{f}=f m_0$, and $\hat{a}=a/w_0$. 

\section{Numerical results: meson masses and decay constants}

\begin{table}
\begin{center}
\begin{tabular}{|c|c|c|c|c|}
\hline\hline
{\rm ~~~Label~}($M$) & {\rm ~Interpolating operator ($\mathcal{O}_M$)~} & $Sp(4)$ & {\rm ~~~$J^{P}$~~~}
& {\rm ~Meson in QCD~}\cr
\hline
PS & $\overline{Q^i}\gamma_5 Q^j$ & $5$ & $0^{-}$ & $\pi$\cr
S & $\overline{Q^i} Q^j$ & $5$ & $0^{+}$ & $a_0$\cr
V & $\overline{Q^i}\gamma_\mu Q^j$ & $10$ & $1^{-}$ & $\rho$\cr
T & $\overline{Q^i}\gamma_0\gamma_\mu Q^j$ & $10$ & $1^{-}$  & $\rho$ \cr
AV & $\overline{Q^i}\gamma_5\gamma_\mu Q^j$ & $5$ & $1^{+}$ & $a_1$ \cr
AT & $\overline{Q^i}\gamma_5\gamma_0\gamma_\mu Q^j$ & $10$ & $1^{+}$ & $b_1$ \cr
\hline\hline
\end{tabular}
\end{center}
\caption{
Interpolating operators sourcing the lightest spin-0 and spin-1 mesons. 
Colour and spinor indices are implicitly summed over, 
while flavour non-singlet mesons are selected by the choices $i\neq j$. 
Spin and parity quantum numbers are denoted by $J^P$ 
(charge conjugation is trivial). 
We show also the irreducible representations of the unbroken global $Sp(4)$ 
and the corresponding mesons in QCD, for comparison. 
}
\label{tab:interpolating_ops}
\end{table}

Our primary goal is to calculate the masses and decay constants of spin-0 and spin-1 mesons. 
To do so, we recall the standard procedure: we first construct  meson 2-point correlation functions 
at positive Euclidean time $\tau$ and zero momentum, by
 using the interpolating operators $\mathcal{O}_M$ listed in Table~\ref{tab:interpolating_ops}.
We then compute the correlators defined as $C_{\mathcal{O}_M}(\tau)\equiv \sum_{\vec{x}} 
\langle  0| \mathcal{O}_{M}(\vec{x},t) \mathcal{O}_M^\dagger(\vec{0},0) | 0 \rangle$. 
At sufficiently large Euclidean time the correlation functions 
are saturated by the lightest states and decay as a single exponential.
The decay rates are identified with the meson masses, while the vacuum-to-meson matrix elements 
containing V and AV currents yield the decay constants of PS, V and AV mesons. 
Matrix elements computed with Wilson fermions receive multiplicative renormalisation.
We renormalise the decay constants via one-loop perturbative matching
 with tadpole improvements for the gauge coupling. 
(For more details see Ref.~\cite{Bennett:2019jzz}.) 

\begin{figure}[t]
\begin{center}
\includegraphics[width=.43\textwidth]{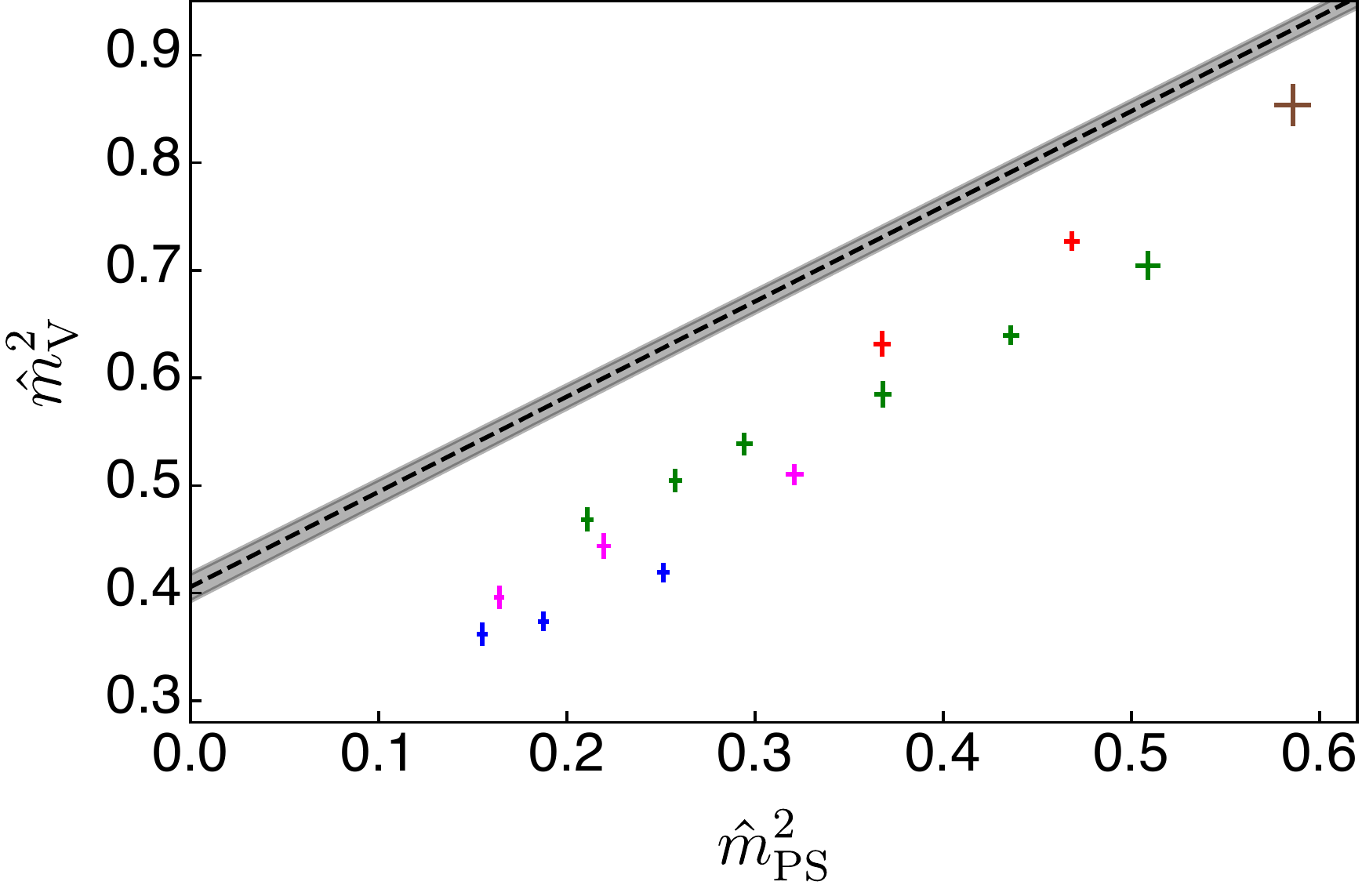}
\includegraphics[width=.43\textwidth]{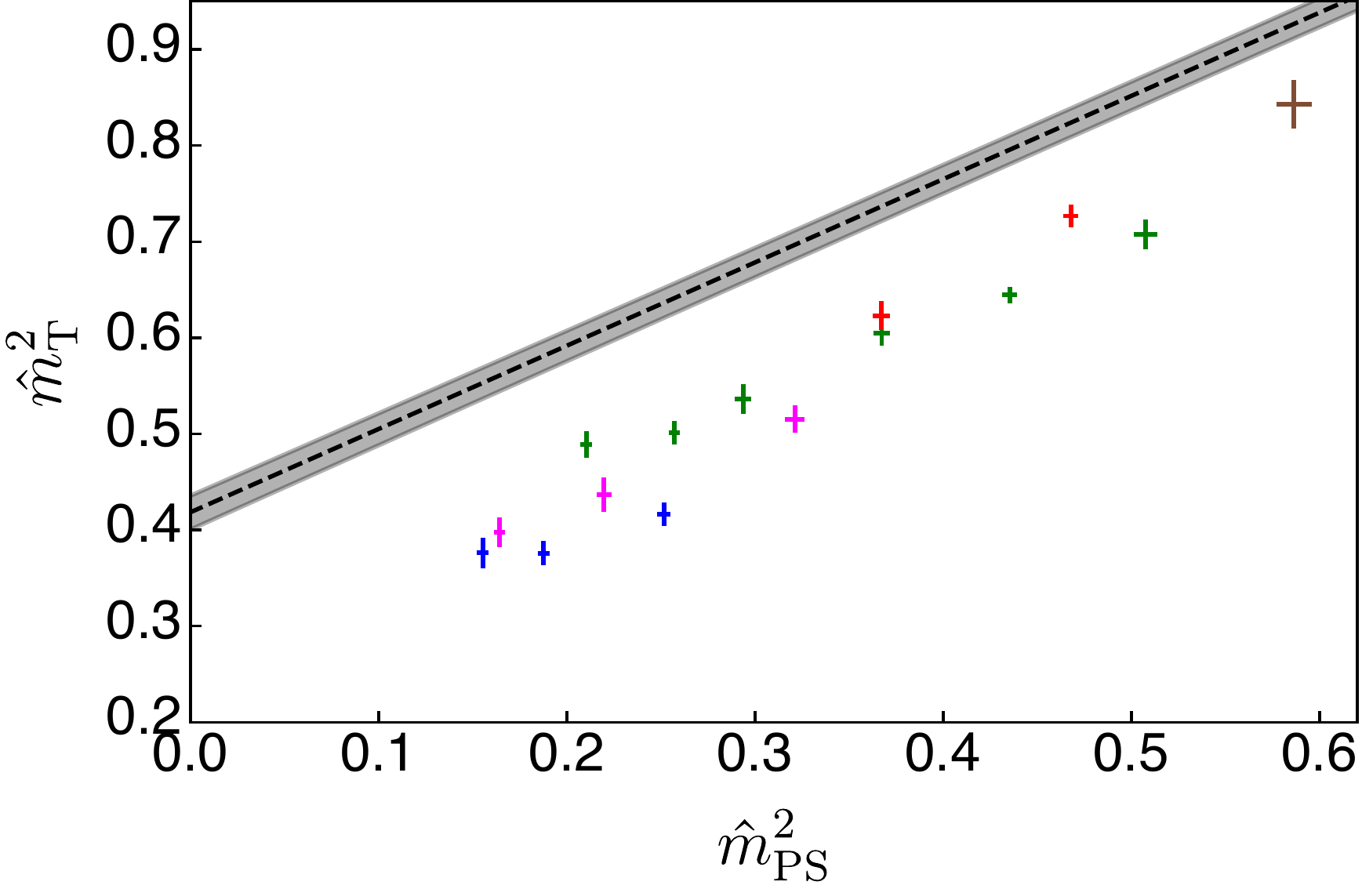}
\includegraphics[width=.43\textwidth]{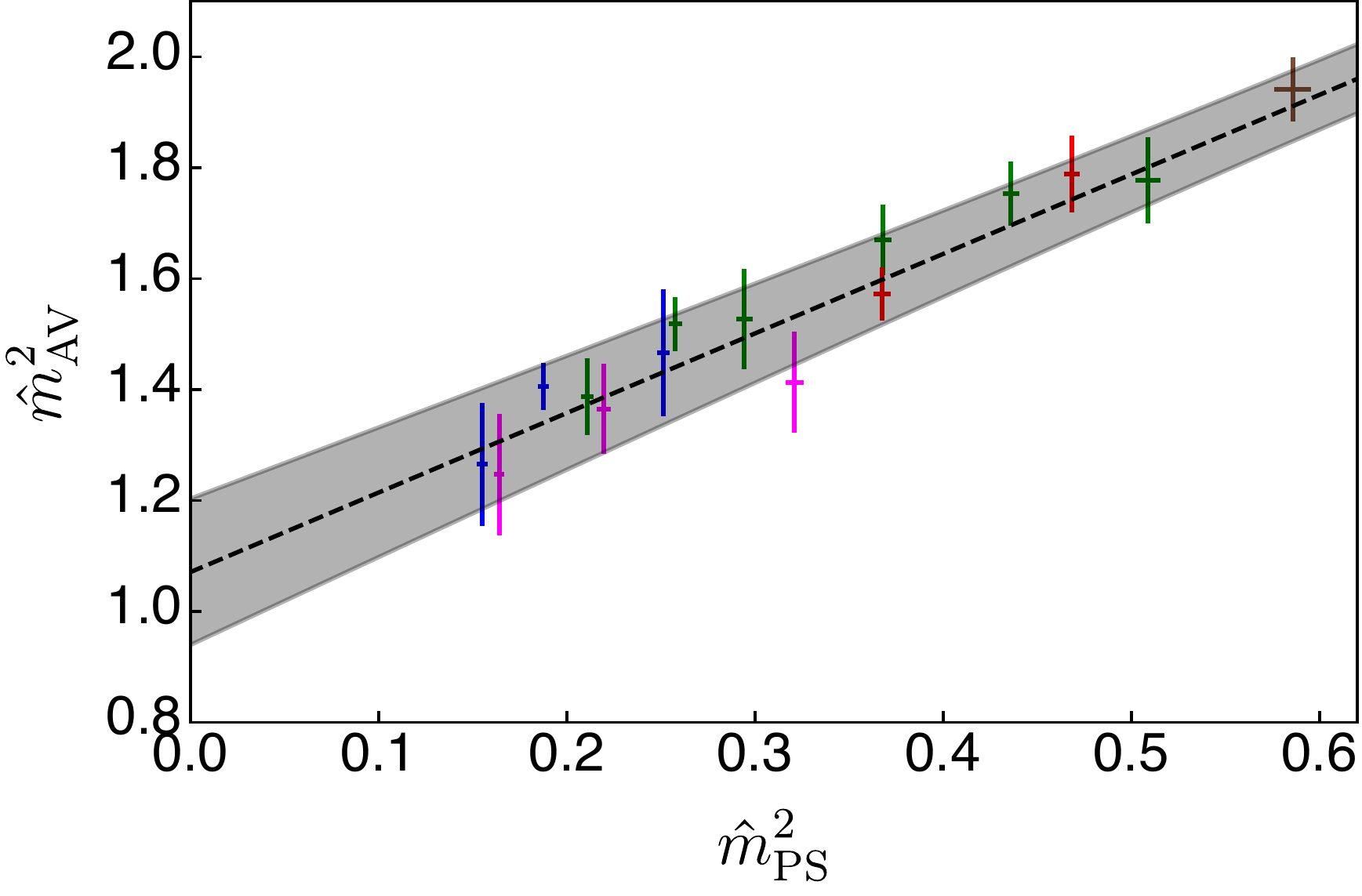}
\includegraphics[width=.43\textwidth]{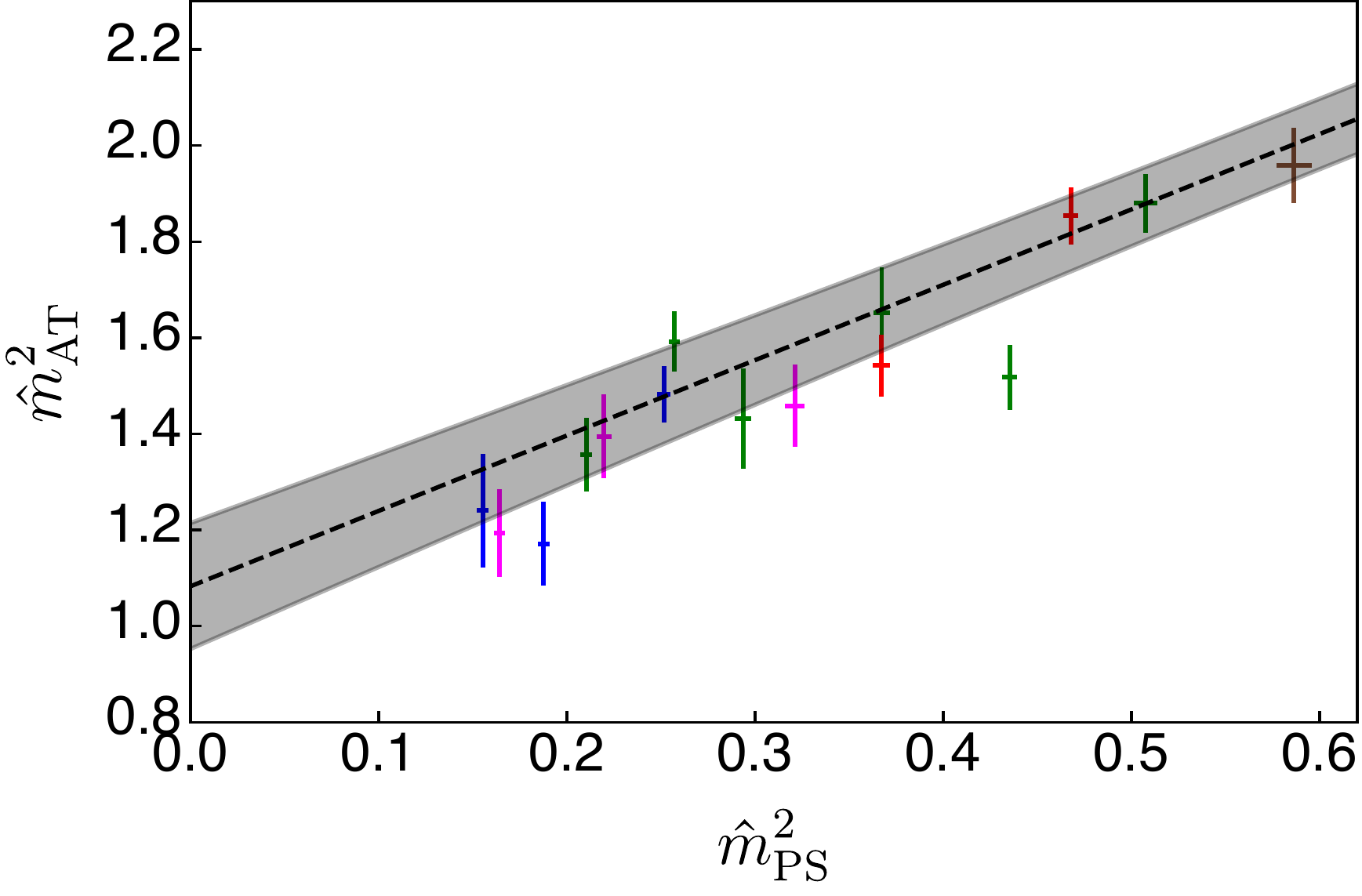}
\includegraphics[width=.43\textwidth]{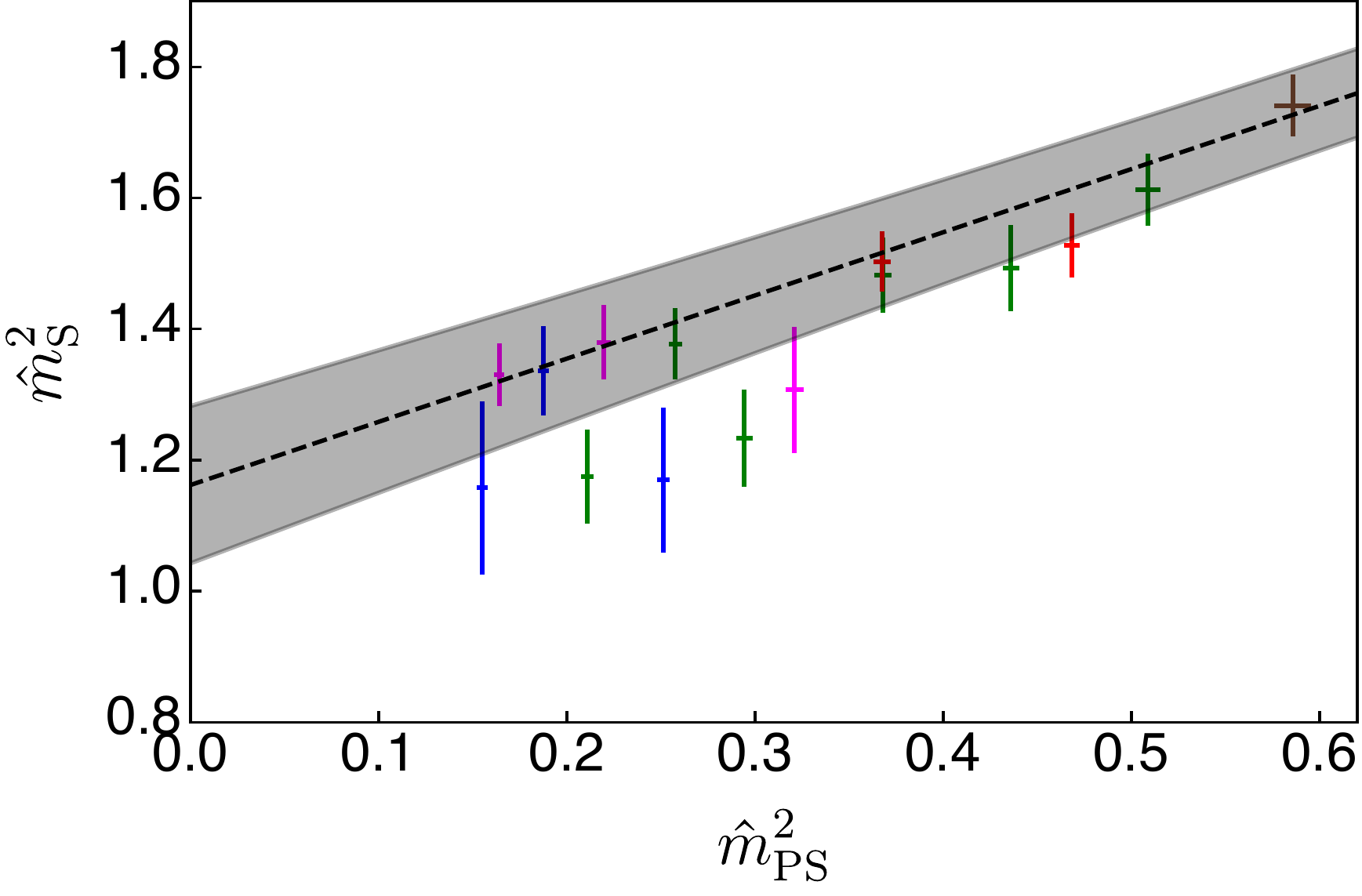}
\caption{%
\label{fig:m2_NLO_fit}%
Masses squared of the vector (V), 
tensor (T), axial-vector (AV), axial-tensor (AT), and scalar (S) mesons, 
as a function of the pseudoscalar (PS) mass squared $\hat{m}_{\rm PS}^2$. 
Blue, purple, green, red and brown colours represent different lattice couplings: $\beta=6.9$ 
(blue), $7.05$ (purple), $7.2$ (green), $7.4$ (red), and $7.5$ (brown). 
The fit results for the continuum and massless extrapolations, 
after subtracting discretisation artefacts, are denoted by the grey bands, 
the widths of which represent the statistical uncertainties. 
}
\end{center}
\end{figure}

\begin{figure}[t]
\begin{center}
\includegraphics[width=.43\textwidth]{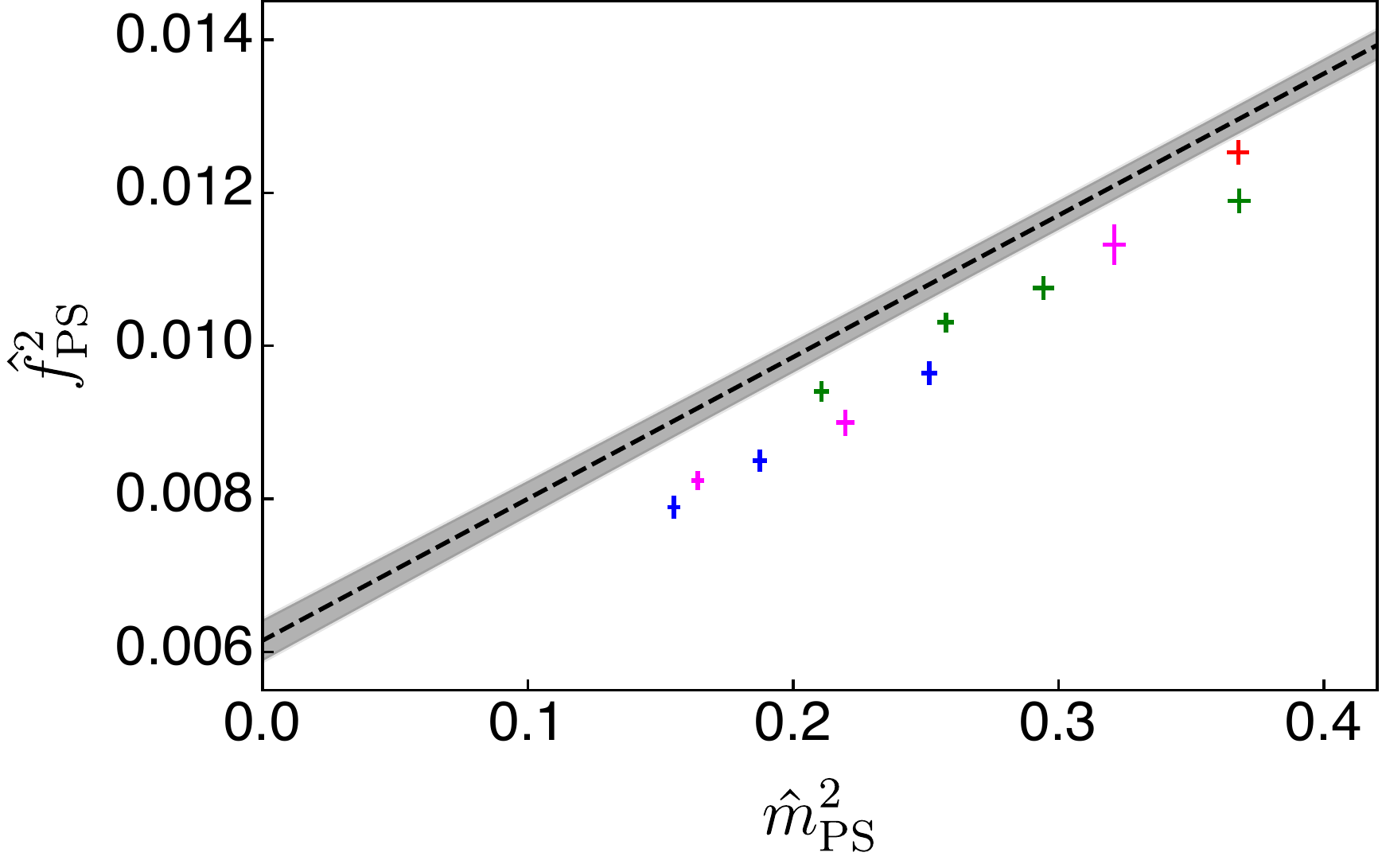}
\includegraphics[width=.43\textwidth]{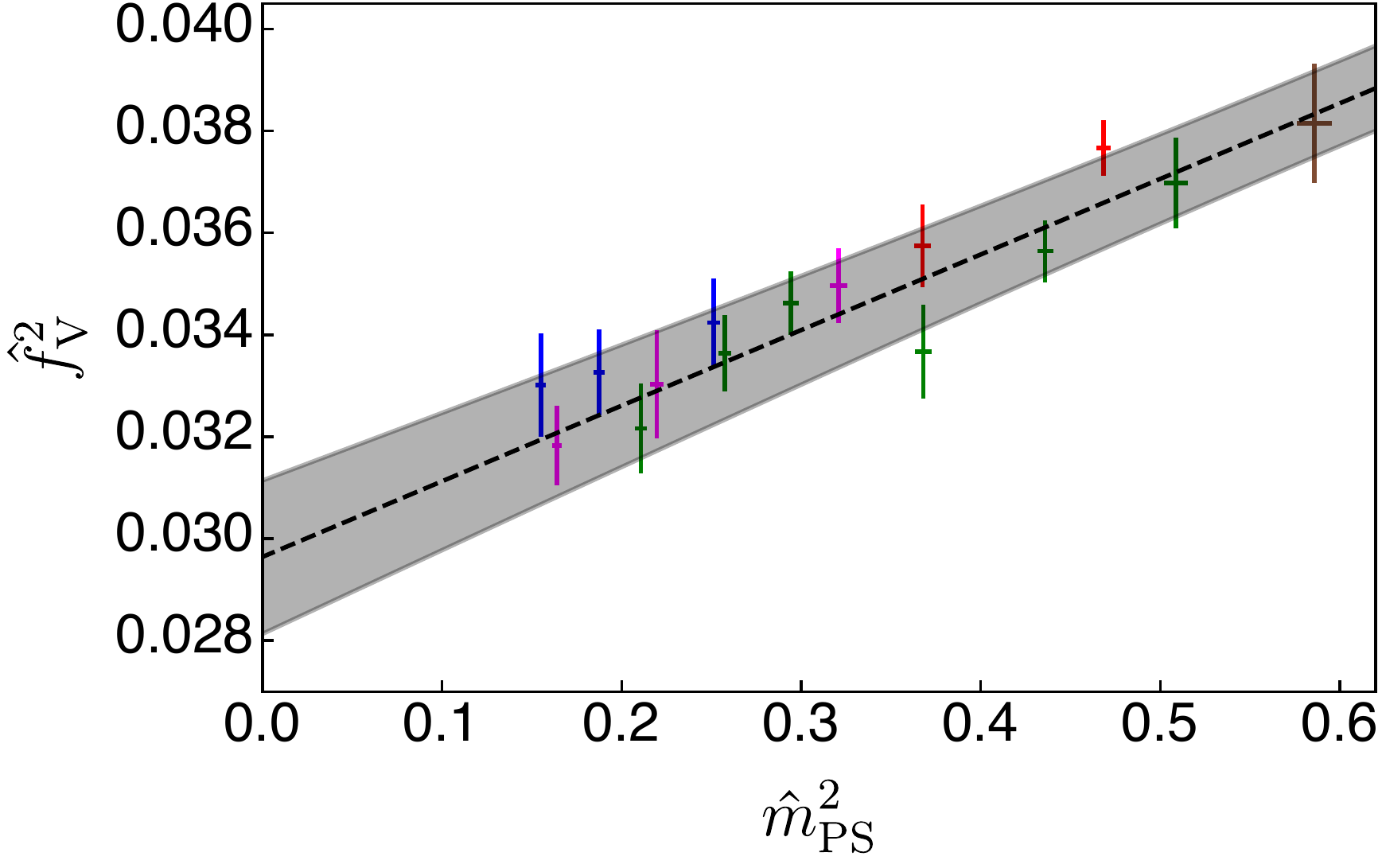}
\includegraphics[width=.43\textwidth]{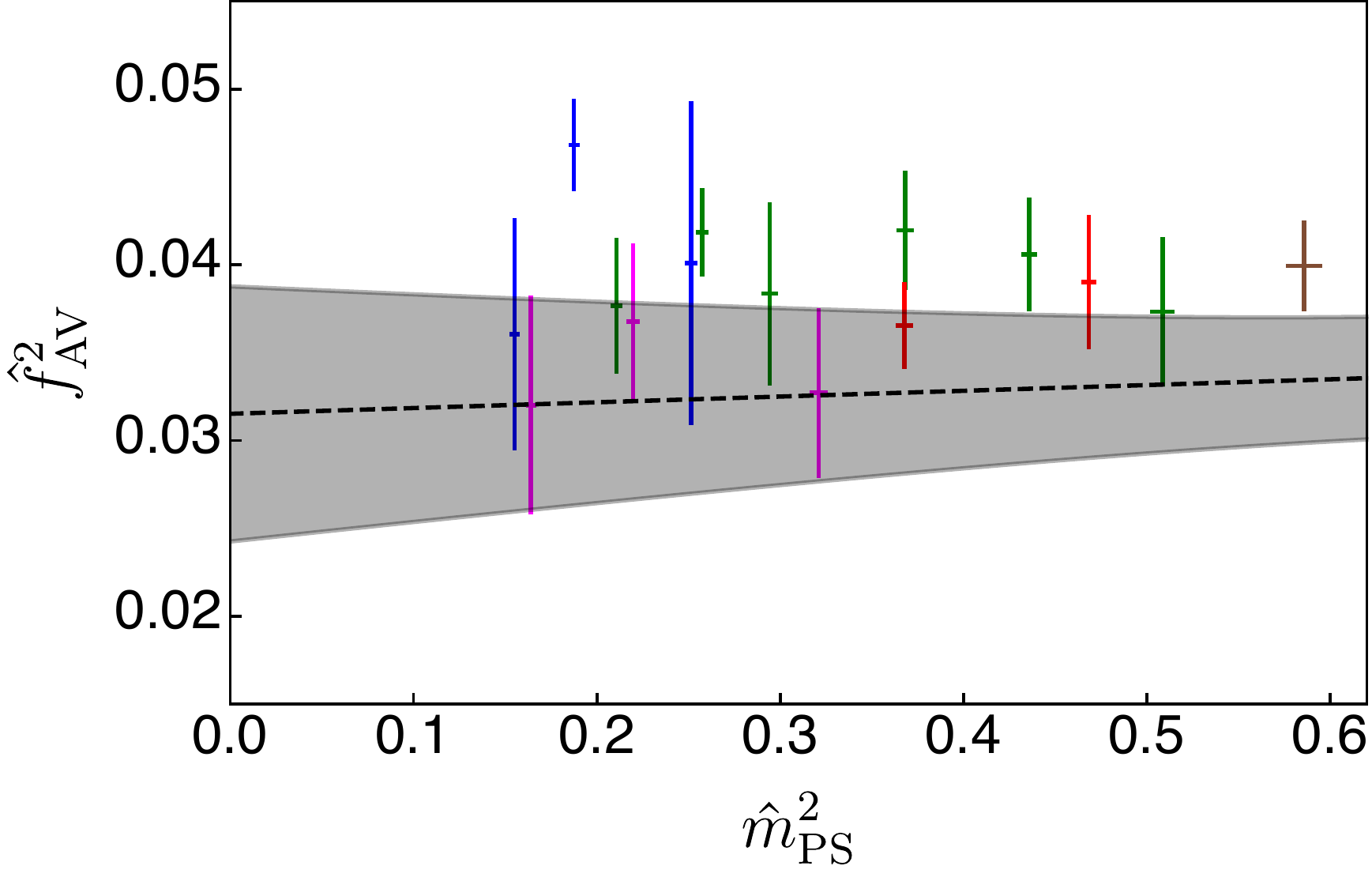}
\caption{%
\label{fig:f2_NLO_fit}%
Decay constants squared of the 
pseudoscalar (PS), vector (V), and axial-vector (AV) mesons, 
as a function of the pseudoscalar (PS) mass squared $\hat{m}_{\rm PS}^2$,
following the same conventions as in Fig.~\ref{fig:m2_NLO_fit}. 
}
\end{center}
\end{figure}

The scale-setting procedure discussed earlier on allows us to 
treat all the dimensional quantities measured at different lattice couplings in a consistent manner, 
by eliminating the explicit dependence on the coupling. 
Residual corrections are due to discretisation effects. 
Furthermore, the moderate to large values of the fermion mass considered 
are such as to render the V mesons stable. 
To remove  lattice artefacts and access the small-mass regime, 
we use the following linear ans\"{a}tz for the mass squared and decay constant squared 
of the mesons, inspired by tree-level next-to-leading-order (NLO) W$\chi$PT:
\begin{eqnarray}
\label{eq:f_chipt}
\hat{f}_M^{2,{\rm NLO}}&=&\hat{f}_M^{2, \chi}\left(1 + L_{f,M}^0 \hat{m}_{\rm PS}^2\right) + W_{f,M}^0 \hat{a},\\
\hat{m}_M^{2,{\rm NLO}}&=&\hat{m}_M^{2, \chi}\left(1 + L_{m,M}^0 \hat{m}_{\rm PS}^2\right) + W_{m,M}^0 \hat{a}. 
\label{eq:m_chipt}
\end{eqnarray}
Notice that we replace the fermion mass in the original W$\chi$PT 
with the pseudoscalar mass squared, 
by making use of the leading order $\chi$PT relation  $m_{\rm PS}^2 = 2 B m_f$. 
We find that these linear ans\"{a}tzs describe the data well up to $m_{\rm PS}^2 \lesssim 0.4$ for the pseudoscalar, 
and $m_{\rm PS}^2 \lesssim 0.6$ for all other mesons. 
The lattice spacing satisfies $\hat{a} \lesssim 1.0$. 
The numerical results are shown in Fig.~\ref{fig:m2_NLO_fit} for the mass squared 
and in Fig.~\ref{fig:f2_NLO_fit} for the decay constant squared.
Different colours denote  different lattice couplings and the grey bands represent the fit results 
after subtracting the lattice artefacts. 
The quantities $\hat{f}_{\rm PS}^2$, $\hat{m}_{\rm V}^2$ and $\hat{m}_{\rm T}^2$ 
are affected by substantial finite lattice spacing corrections. 
The agreement between V and T masses is consistent with the fact that, 
the global symmetry being broken, 
these two states mix. 

\section{Low-energy EFT and phenomenology}

\begin{figure}[t]
\begin{center}
\includegraphics[width=.48\textwidth]{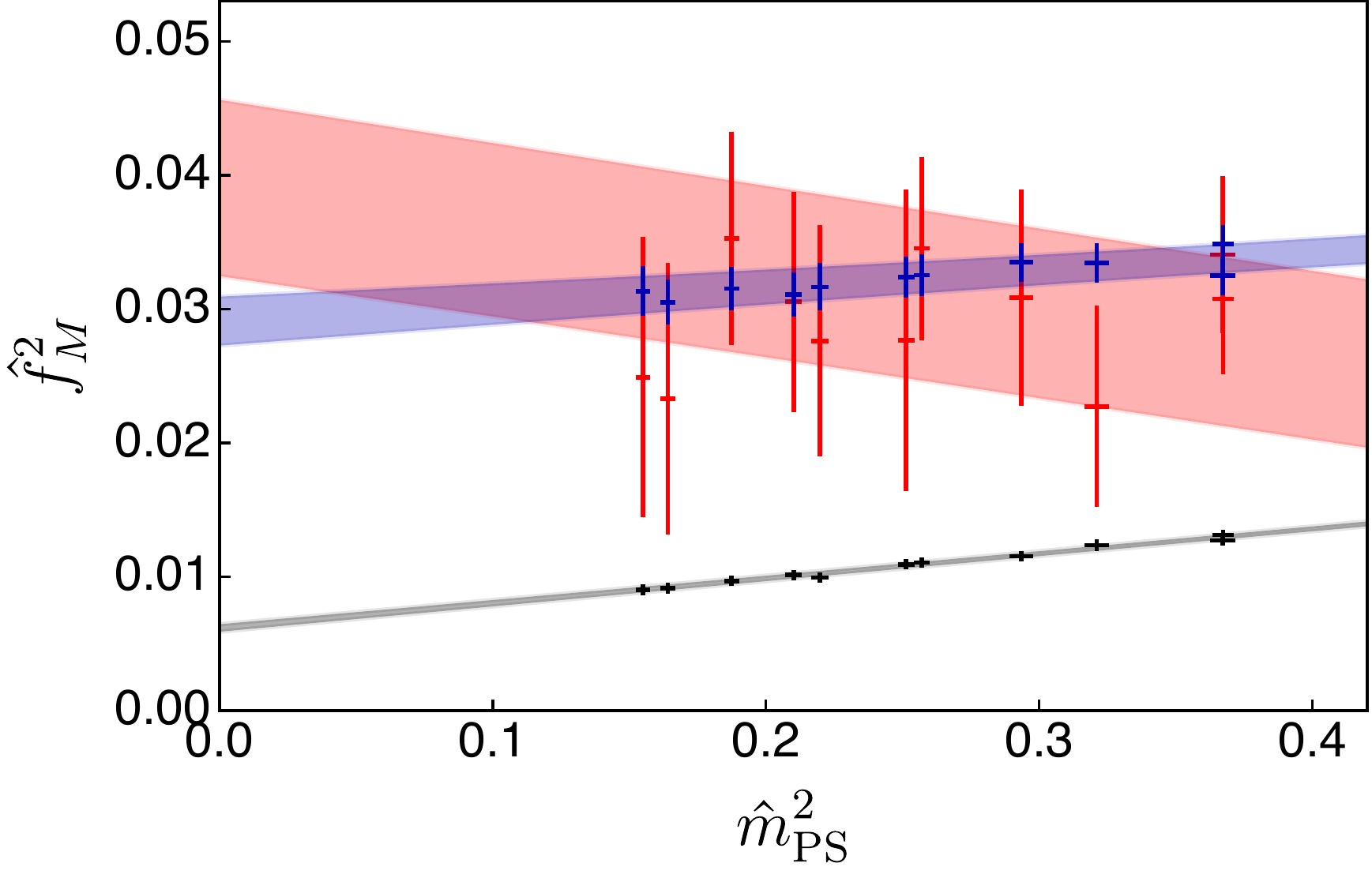}
\includegraphics[width=.47\textwidth]{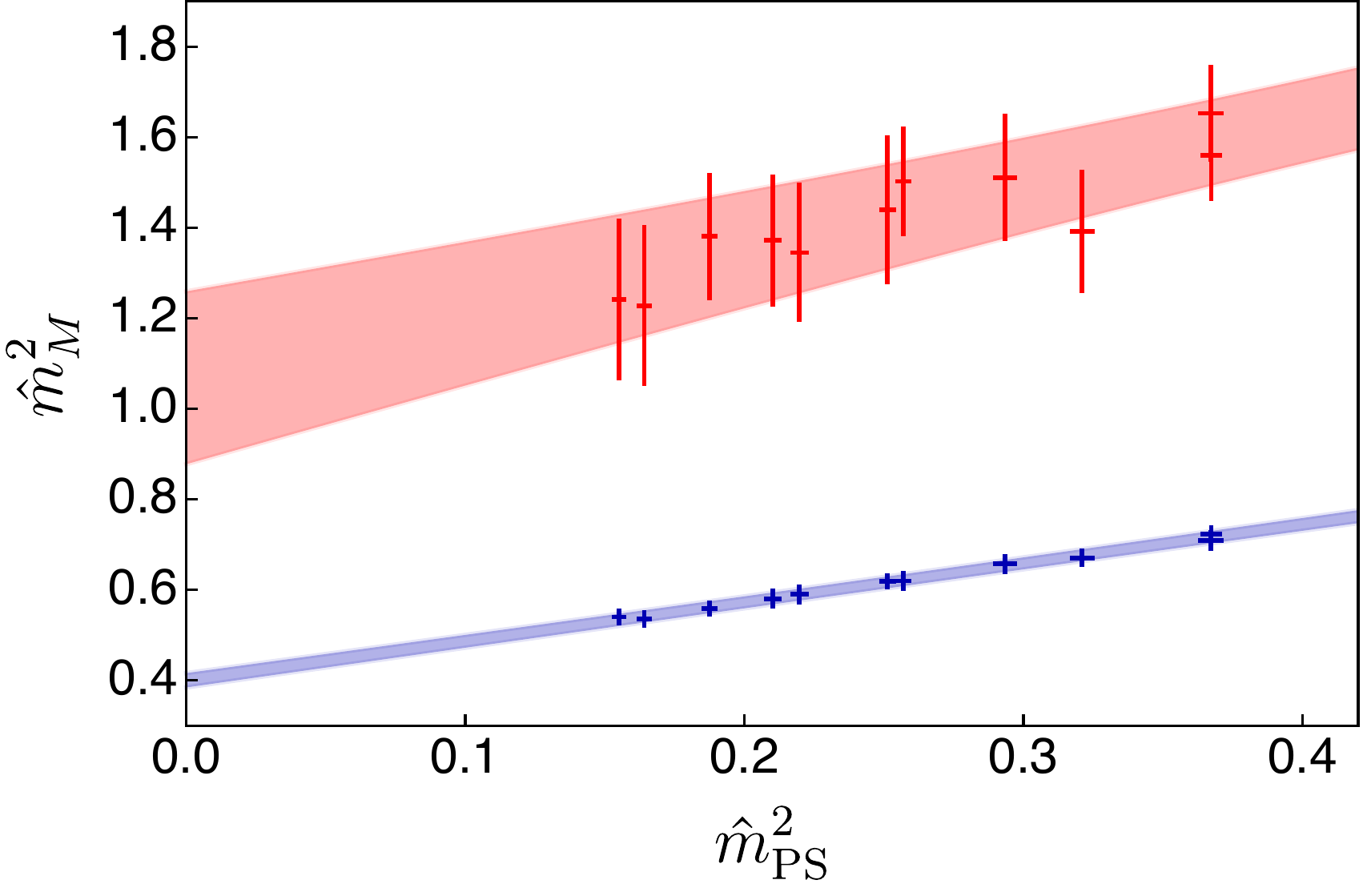}
\caption{%
\label{fig:gfit}%
Continuum-extrapolated meson masses and decay constants squared 
as a function of the 
pseudoscalar mass squared. 
Black, blue and red colours are for PS, V and AV mesons. 
The global fit results are denoted by shaded bands with their widths representing the statistical errors.
}
\end{center}
\end{figure}

The chiral Lagrangian, the effective theory describing the long distance behaviour of PS mesons, 
can be extended to describe the lightest V and AV mesons
by adopting the ideas of hidden local symmetry (HLS)~\cite{Bando:1984ej}. 
We apply this idea to the two-flavour $Sp(4)$ theory 
and construct the tree-level NLO HLS effective Lagrangian in Ref.~\cite{Bennett:2017kga}. 
Meson masses and decay constants can be expressed in terms of 
the low-energy constants (LECs) in such effective Lagrangian. 
Since we replace the fermion mass by the PS mass squared, after confirming the linear behaviour of the observables, 
we rewrite the original expressions in Ref.~\cite{Bennett:2017kga} 
by only keeping the terms up to the leading order in an expansion in $m_{\rm PS}^2$. 
Using the final expressions shown in Eqs.~(6.1)-(6.5) of Ref.~\cite{Bennett:2019jzz}, 
we perform the global fit involving $10$ LECs to the continuum-extrapolated results of 
$\hat{f}_{\rm PS}^2$, $\hat{f}_{\rm V}^2$, $\hat{f}_{\rm AV}^2$, $\hat{m}_{\rm V}^2$,
 and $\hat{m}_{\rm AV}^2$. 
As shown by Fig.~\ref{fig:gfit}, the EFT describes 
the numerical results well, as is also indicated by the result of the fit itself
at the minimum of $\chi^2$, for which we find that $\chi^2/N_{\rm d.o.f}\sim 0.4$.

The coupling $g_{\rm VPP}$ controls the decay of one V  into two PS states, 
and appears within the HLS EFT~\cite{Bennett:2017kga}. 
Using the results from the global fit, in the massless limit,
 we extrapolate the coupling $g_{\rm VPP}^\chi=6.0(4)(2)$,
where in parenthesis we report the statistical and systematic errors, respectively. 
The former corresponds the $1\sigma$ standard deviation, 
while the latter is estimated by varying the fitting range to include/exclude the coarsest/heaviest ensemble. 
A phenomenological relation, 
based on the vector meson dominance, 
was proposed by Kawarabayashi, Suzuki, Riazuddin and Fayyazuddin (KSRF) 
stating that $g_{\rm VPP}=m_{\rm V}/\sqrt{2} f_{\rm PS}$ \cite{Kawarabayashi:1966kd}. 
We find that our result for $\hat{m}_{\rm V}/\sqrt{2} \hat{f}_{\rm PS}=5.72(18)(13)$,
extrapolated to the massless limit, 
is in good agreement with the coupling estimated from the EFT, 
providing some empirical  support to the KSRF relation. 

\begin{figure}[t]
\begin{center}
\includegraphics[width=.56\textwidth]{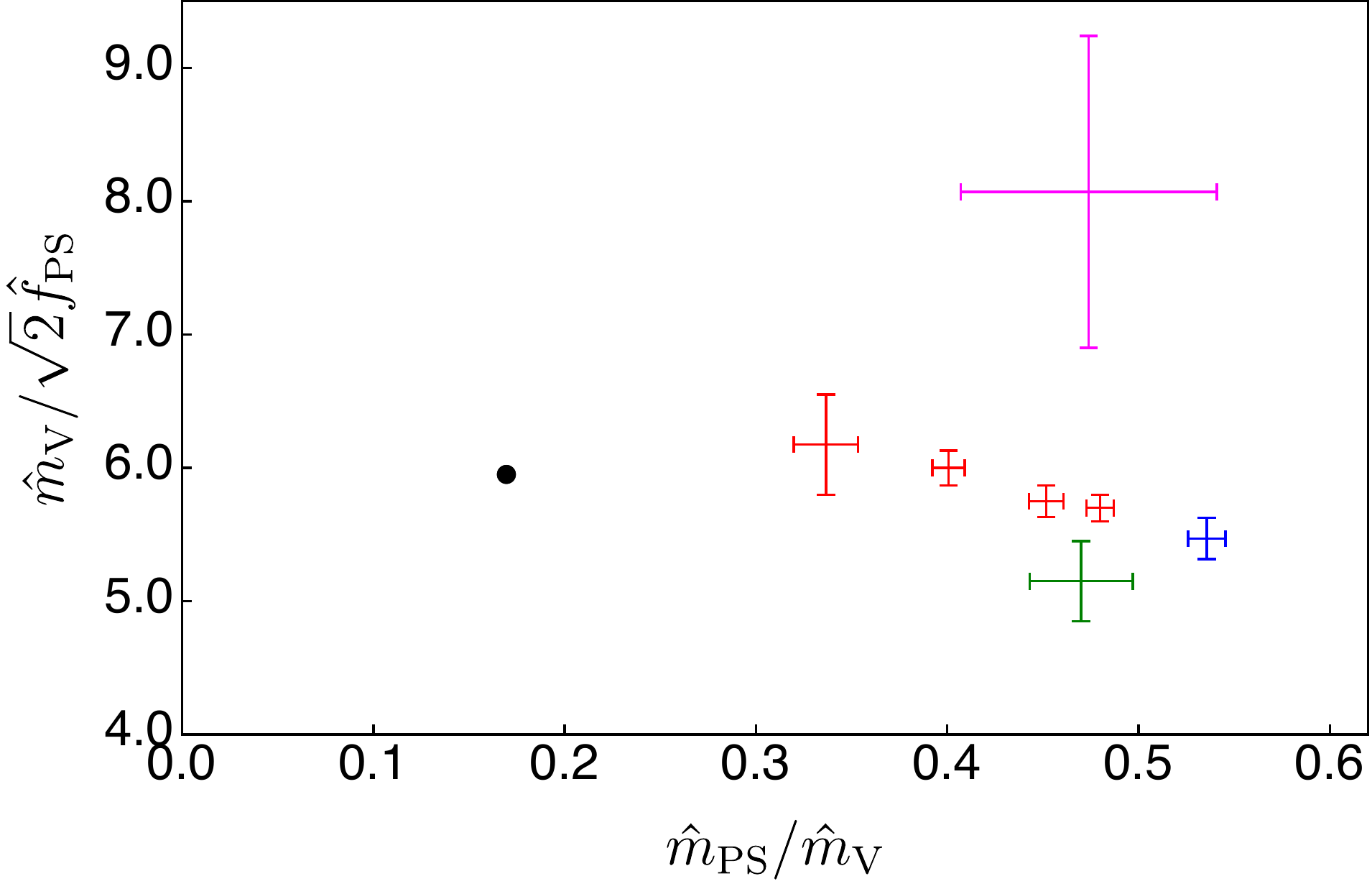}
\caption{%
\label{fig:gvpp_comparison}%
Comparison between the ratios $m_{\rm V}/\sqrt{2}f_{\rm PS}$
 computed from several different lattice gauge theories 
with $N_f=2$ fundamental Dirac fermions. 
Purple, red, green, and blue colours denote 
the $SU(2)$~\cite{Arthur:2016dir}, $SU(3)$~\cite{Jansen:2009hr}, 
$SU(4)$~\cite{Ayyar:2017qdf}, and $Sp(4)$ gauge group, respectively. 
For reference, we also show the experimental value of the coupling in real-world QCD, 
denoted by a black dot. 
}
\end{center}
\end{figure}

In Fig. \ref{fig:gvpp_comparison}, 
we make a comparison of the result of this ratio estracted from our lightest ensemble
with  lattice results 
obtained in other theories with two fundamental Dirac flavours: in $SU(2)$~\cite{Arthur:2016dir}, 
in $SU(3)$~\cite{Jansen:2009hr}, and in $SU(4)$~\cite{Ayyar:2017qdf}. 
The $N$ dependence of the ratio is consistent with large-$N$
 arguments, according to which one expects $m_{\rm V}\sim\textrm{constant}$ and $f_{\rm PS}\sim \sqrt{N}$. 

\section{Conclusion}
We performed numerical studies of some of the lightest spin-0 and spin-1 flavoured mesons 
in  the $Sp(4)$ lattice gauge theory with $N_f=2$ dynamical Dirac fermions. 
From several ensembles characterised by different lattice couplings and fermion masses,
we carried out continuum extrapolations, using Eqs.~(\ref{eq:f_chipt}) and~(\ref{eq:m_chipt}). 
We also analysed the continuum-extrapolated data by making use of a low-energy EFT description
that allowed us to discuss phenomenological implications such as the coupling
between one V and two PS states. 
The extrapolated results provide access to the small-mass regime of relevance in the context of 
composite Higgs models, while the large-mass regime directly investigated is relevant 
in the context of phenomenological studies of SIMPs. 

\begin{acknowledgments}
We would like to thank Michele Mesiti and Jarno Rantaharju for their assistance on 
the modification and improvement of the HiRep code for this project. 

\end{acknowledgments}


\begin{thebibliography}{99}

\bibitem{Kaplan:1983fs} 
  D.~B.~Kaplan and H.~Georgi,
  Phys.\ Lett.\  {\bf 136B}, 183 (1984); 
  Phys.\ Lett.\  {\bf 145B}, 216 (1984).

\bibitem{Kaplan:1991dc} 
  D.~B.~Kaplan,
  Nucl.\ Phys.\ B {\bf 365}, 259 (1991).
  doi:10.1016/S0550-3213(05)80021-5

\bibitem{Bennett:2017kga} 
  E.~Bennett, D.~K.~Hong, J.~W.~Lee, C.-J.~D.~Lin, B.~Lucini, M.~Piai and D.~Vadacchino,
  JHEP {\bf 1803}, 185 (2018)
  [arXiv:1712.04220 [hep-lat]].
 
\bibitem{Hochberg:2014kqa} 
  Y.~Hochberg, E.~Kuflik, H.~Murayama, T.~Volansky and J.~G.~Wacker,
  Phys.\ Rev.\ Lett.\  {\bf 115}, no. 2, 021301 (2015)
  [arXiv:1411.3727 [hep-ph]].

\bibitem{Bennett:2019jzz} 
E.~Bennett, D.~K.~Hong, J.~W.~Lee, C.-J.~D.~Lin, B.~Lucini, M.~Piai and D.~Vadacchino,
  arXiv:1909.12662 [hep-lat].

\bibitem{quenched}
  E.~Bennett, D.~K.~Hong, J.~W.~Lee, C.-J.~D.~Lin, B.~Lucini, M.~Mesiti, M.~Piai, J.~Rantaharju and D.~Vadacchino,
  in preparation.
 
\bibitem{DelDebbio:2008zf} 
  L.~Del Debbio, A.~Patella and C.~Pica,
  Phys.\ Rev.\ D {\bf 81}, 094503 (2010)
  [arXiv:0805.2058 [hep-lat]].
 
\bibitem{Lee:2018ztv} 
  J.~W.~Lee, E.~Bennett, D.~K.~Hong, C.~J.~D.~Lin, B.~Lucini, M.~Piai and D.~Vadacchino,
  PoS LATTICE {\bf 2018}, 192 (2018)
  [arXiv:1811.00276 [hep-lat]].
 
\bibitem{Luscher:2010iy} 
  M.~L\"{u}scher,
  JHEP {\bf 1008}, 071 (2010)
  Erratum: [JHEP {\bf 1403}, 092 (2014)]
  [arXiv:1006.4518 [hep-lat]].
 
\bibitem{Borsanyi:2012zs} 
  S.~Borsanyi {\it et al.},
  JHEP {\bf 1209}, 010 (2012)
  [arXiv:1203.4469 [hep-lat]].
 
\bibitem{Bando:1984ej} 
  M.~Bando, T.~Kugo, S.~Uehara, K.~Yamawaki and T.~Yanagida,
  Phys.\ Rev.\ Lett.\  {\bf 54}, 1215 (1985).

\bibitem{Kawarabayashi:1966kd} 
  K.~Kawarabayashi and M.~Suzuki,
  Phys.\ Rev.\ Lett.\  {\bf 16}, 255 (1966);
  Riazuddin and Fayyazuddin,
  Phys.\ Rev.\  {\bf 147}, 1071 (1966).

\bibitem{Arthur:2016dir} 
  R.~Arthur, V.~Drach, M.~Hansen, A.~Hietanen, C.~Pica and F.~Sannino,
  Phys.\ Rev.\ D {\bf 94}, no. 9, 094507 (2016)
  [arXiv:1602.06559 [hep-lat]].

\bibitem{Jansen:2009hr} 
  K.~Jansen {\it et al.} [ETM Collaboration],
  Phys.\ Rev.\ D {\bf 80}, 054510 (2009)
  [arXiv:0906.4720 [hep-lat]].
 
\bibitem{Ayyar:2017qdf} 
  V.~Ayyar, T.~DeGrand, M.~Golterman, D.~C.~Hackett, W.~I.~Jay, E.~T.~Neil, Y.~Shamir and B.~Svetitsky,
  Phys.\ Rev.\ D {\bf 97}, no. 7, 074505 (2018)
  [arXiv:1710.00806 [hep-lat]].

\end{thebibliography}
\end{document}